\documentstyle[preprint,tighten,aps]{revtex}
\begin{document}
\input{psfig}
\draft
\title{Limits on $WWZ$ and $WW\gamma$ couplings from
$p\bar{p}\rightarrow e\nu jj X$ events at $\sqrt{s} = 1.8$ TeV}
\preprint{Fermilab--Pub--97/136--E}

%
\author{                                                                      
B.~Abbott,$^{28}$                                                             
M.~Abolins,$^{25}$                                                            
B.S.~Acharya,$^{43}$                                                          
I.~Adam,$^{12}$                                                               
D.L.~Adams,$^{37}$                                                            
M.~Adams,$^{17}$                                                              
S.~Ahn,$^{14}$                                                                
H.~Aihara,$^{22}$                                                             
G.A.~Alves,$^{10}$                                                            
E.~Amidi,$^{29}$
% ...
\\                                                                            
\vskip 0.50cm                                                                 
\centerline{(D\O\ Collaboration)}                                             
\vskip 0.50cm                                                                 
}                                                                             
\address{                                                                     
\centerline{$^{1}$Universidad de los Andes, Bogot\'{a}, Colombia}             
\centerline{$^{2}$University of Arizona, Tucson, Arizona 85721}               
\centerline{$^{3}$Boston University, Boston, Massachusetts 02215}             
\centerline{$^{4}$Brookhaven National Laboratory, Upton, New York 11973}      
\centerline{$^{5}$Brown University, Providence, Rhode Island 02912}           
\centerline{$^{6}$Universidad de Buenos Aires, Buenos Aires, Argentina}       
\centerline{$^{7}$University of California, Davis, California 95616}          
\centerline{$^{8}$University of California, Irvine, California 92697}         
\centerline{$^{9}$University of California, Riverside, California 92521}      
\centerline{$^{10}$LAFEX, Centro Brasileiro de Pesquisas F{\'\i}sicas,        
                  Rio de Janeiro, Brazil}                                     
\centerline{$^{11}$CINVESTAV, Mexico City, Mexico}                            
\centerline{$^{12}$Columbia University, New York, New York 10027}             
\centerline{$^{13}$Delhi University, Delhi, India 110007}                     
\centerline{$^{14}$Fermi National Accelerator Laboratory, Batavia,            
                   Illinois 60510}                                            
\centerline{$^{15}$Florida State University, Tallahassee, Florida 32306}      
\centerline{$^{16}$University of Hawaii, Honolulu, Hawaii 96822}              
\centerline{$^{17}$University of Illinois at Chicago, Chicago,                
                   Illinois 60607}                                            
\centerline{$^{18}$Indiana University, Bloomington, Indiana 47405}            
\centerline{$^{19}$Iowa State University, Ames, Iowa 50011}                   
\centerline{$^{20}$Korea University, Seoul, Korea}                            
\centerline{$^{21}$Kyungsung University, Pusan, Korea}                        
\centerline{$^{22}$Lawrence Berkeley National Laboratory and University of    
                   California, Berkeley, California 94720}                    
\centerline{$^{23}$University of Maryland, College Park, Maryland 20742}      
\centerline{$^{24}$University of Michigan, Ann Arbor, Michigan 48109}         
\centerline{$^{25}$Michigan State University, East Lansing, Michigan 48824}   
\centerline{$^{26}$Moscow State University, Moscow, Russia}                   
\centerline{$^{27}$University of Nebraska, Lincoln, Nebraska 68588}           
\centerline{$^{28}$New York University, New York, New York 10003}             
\centerline{$^{29}$Northeastern University, Boston, Massachusetts 02115}      
\centerline{$^{30}$Northern Illinois University, DeKalb, Illinois 60115}      
\centerline{$^{31}$Northwestern University, Evanston, Illinois 60208}         
\centerline{$^{32}$University of Notre Dame, Notre Dame, Indiana 46556}       
\centerline{$^{33}$University of Oklahoma, Norman, Oklahoma 73019}            
\centerline{$^{34}$University of Panjab, Chandigarh 16-00-14, India}          
\centerline{$^{35}$Institute for High Energy Physics, 142-284 Protvino,       
                   Russia}                                                    
\centerline{$^{36}$Purdue University, West Lafayette, Indiana 47907}          
\centerline{$^{37}$Rice University, Houston, Texas 77005}                     
\centerline{$^{38}$Universidade Estadual do Rio de Janeiro, Brazil}           
\centerline{$^{39}$University of Rochester, Rochester, New York 14627}        
\centerline{$^{40}$CEA, DAPNIA/Service de Physique des Particules,            
                   CE-SACLAY, Gif-sur-Yvette, France}                         
\centerline{$^{41}$Seoul National University, Seoul, Korea}                   
\centerline{$^{42}$State University of New York, Stony Brook,                 
                   New York 11794}                                            
\centerline{$^{43}$Tata Institute of Fundamental Research,                    
                   Colaba, Mumbai 400005, India}                              
\centerline{$^{44}$University of Texas, Arlington, Texas 76019}               
\centerline{$^{45}$Texas A\&M University, College Station, Texas 77843}       
}                                                                             

\maketitle
\vspace{-0.2in}
\begin{abstract}

We  present limits  on  anomalous $WWZ$  and  $WW\gamma$  couplings from a
search  for  $WW$  and  $WZ$   production  in   $p\bar{p}$   collisions at
$\sqrt{s}=1.8$ TeV. We use  $p\bar{p}\rightarrow e\nu jjX$ events recorded
with the  D\O\  detector  at the  Fermilab  Tevatron  Collider  during the
1992--1995 run. The data sample corresponds to an integrated luminosity of
$96.0\pm 5.1$~pb$^{-1}$. Assuming  identical $WWZ$ and $WW\gamma$ coupling
parameters,  the  95\% CL  limits on the   $CP$--conserving  couplings are
$-0.33<\lambda<0.36$    ($\Delta\kappa=0$) and   $-0.43<\Delta\kappa<0.59$
($\lambda=0$), for a  form factor scale $\Lambda  = 2.0$ TeV. Limits based
on other assumptions are also presented.

\end{abstract}

\pacs{\it Submitted to Phys. Rev. Lett.}

The vector boson  trilinear  couplings predicted by  the non-Abelian gauge
symmetry  of the  Standard  Model (SM)  can be  measured  directly in pair
production      processes   such  as       $q\bar{q}\rightarrow   W^+W^-$,
$W^{\pm}\gamma$,    $Z\gamma$,  and  $W^{\pm}Z$.   Deviations  from the SM
couplings  would signal  new  physics. Studies  of such  effects have been
reported      by   the         UA2~\cite{UA2},         CDF~\cite{CDF}  and
D\O~\cite{one,two,TopPRD}  collaborations. In this letter we report on the
measurement  of  $WWV$  couplings (where  $V =  \gamma$ or  $Z$) using the
diboson production processes $p\bar{p}\rightarrow WWX\rightarrow e\nu jjX$
and $p\bar{p}\rightarrow  WZX\rightarrow e\nu jjX$, where $j$ represents a
jet. 

The Lorentz invariant Lagrangian  which describes the $WW\gamma$ and $WWZ$
interactions  has fourteen  independent  coupling  parameters~\cite{Baur},
seven  describing the  $WW\gamma$ vertex  and seven for  the $WWZ$ vertex.
Assuming   electromagnetic  gauge  invariance and  $CP$  conservation, the
number   of   parameters  is   reduced to   five:   $g_1^Z$,   $\kappa_Z$,
$\kappa_{\gamma}$,  $\lambda_Z$ and  $\lambda_{\gamma}$. In the SM at tree
level, the coupling pa\-ra\-me\-ters  have the values $\Delta g_1^Z(\equiv
g_1^Z    -1)  =   0$,           $\Delta\kappa_Z(\equiv    \kappa_Z   -1) =
\Delta\kappa_{\gamma}(\equiv      \kappa_{\gamma} -1)  = 0$,  $\lambda_Z =
\lambda_{\gamma}  = 0$.  The SM  cross sections  for  $p\bar{p}\rightarrow
W^+W^-X$ and  $p\bar{p}\rightarrow W^{\pm}ZX$  production at the Tevatron,
at $\sqrt{s} =1.8$ TeV, are 9.5~pb and 2.7~pb~\cite{Xsec} respectively. 

Non-SM values of  the coupling  parameters would  result in an increase of
the   production  cross   section,  especially  for  large  values  of the
transverse momentum of the $W$ boson ($p_T^W$). Since tree level unitarity
restricts the  $WWV$ couplings to  their SM values at  asymptotically high
energies,  each of the  couplings must be  modified by a  form factor e.g.
$\lambda_Z(\hat{s}) =  \lambda_Z/(1+\hat{s}/\Lambda^2)^2$, where $\hat{s}$
is the  square  of the   invariant mass  of the  $WW$ or  $WZ$  system and
$\Lambda$ is the form-factor scale. We have used $\Lambda= 1.0$, $1.5$ and
$2.0$ TeV. 

The analysis  reported here  uses  $p\bar{p}\rightarrow  e\nu jj X$ events
recorded  with the  D\O\  detector during  the  1992--1993 and  1993--1995
Fermilab Tevatron Collider runs at  $\sqrt{s} = 1.8$ TeV, corresponding to
a total   integrated   luminosity of  $96.0\pm   5.1$~pb$^{-1}$.  The D\O\
detector       and    data       collection    system    are     described
elsewhere~\cite{Detector}.      The  basic  elements  of  the  trigger and
reconstruction algorithms for jets,  electrons, and neutrinos are given in
Ref.~\cite{TopPRD}. The analysis of $e\nu jj X$ events from the 1992--1993
Tevatron     Collider  run   ($13.7\pm     0.7$~pb$^{-1}$)  was   reported
previously~\cite{two}.     This  letter  focuses  on the   analysis of the
1993--1995  data set of  $82.3\pm   4.4$~pb$^{-1}$ and gives  the combined
results    for  both    analyses.   Further   details  are    available in
Ref.~\cite{Thesis}.

The data  sample was  obtained with a  trigger which  required an isolated
electromagnetic (EM)  calorimeter cluster with  transverse energy $E_T>20$
GeV          and           missing                transverse        energy
${\hbox{$E$\kern-0.6em\lower-.1ex\hbox{/}}}_T > 15$ GeV. The offline event
selection required that the EM  cluster have $|\eta| < 1.1$ in the central
calorimeter or $1.5<|\eta|<2.5$ in an end calorimeter, where $\eta$ is the
pseudorapidity. Electrons were identified by requiring that the EM cluster
pass the shower profile and tracking information criteria, as described in
our earlier  analysis~\cite{two}. The presence  of a neutrino was inferred
from the   ${\hbox{$E$\kern-0.6em\lower-.1ex\hbox{/}}}_T$, calculated from
the vector sum of the $E_T$ measured  in each calorimeter tower. Jets were
reconstructed   using  a  cone  algorithm  with  radius   ${\cal{R}}\equiv
\sqrt{(\Delta\eta)^2+(\Delta\phi)^2}  = 0.5$. To  remove spurious jets due
to detector effects, this analysis used the same quality cuts as were used
in  Ref.~\cite{squarks}. Jets were  required to be  within $|\eta| < 2.5$.
The  jet   energies  were  corrected  for   effects of  jet   energy scale
calibration,      out-of-cone   showering,  energy   from the   underlying
event~\cite{Scale}, and energy loss due to out-of-cone gluon radiation. 

The $WW/WZ$ candidates were selected by searching for events containing an
isolated       electron   with  high    $E^e_T$   ($>  25$    GeV),  large
${\hbox{$E$\kern-0.6em\lower-.1ex\hbox{/}}}_T$   ($> 25$ GeV) and at least
two high $E^j_T$  jets ($> 20$ GeV).  The transverse  mass of the electron
and neutrino  system was  required to be  consistent with  a $W\rightarrow
e\nu$             decay                 ($M_T^{e\nu}   =     [2      E^e_T
{\hbox{$E$\kern-0.6em\lower-.1ex\hbox{/}}}_T(1-\cos   \phi^{e\nu})]^{1/2}>
40$  GeV/c$^2$, where  $\phi^{e\nu}$  is the  azimuthal  angle between the
electron and   ${\hbox{$E$\kern-0.6em\lower-.1ex\hbox{/}}}_T$ vector). The
invariant mass ($m^{jj}$) of the two jet system (the largest in\-va\-riant
mass if there were more  than two jets with  $E^j_T> 20$ GeV in the event)
was  required  to be  $50<  m^{jj} <  110$  GeV/c$^2$,  as  expected for a
$W\rightarrow jj$ or  $Z\rightarrow jj$ decay.  Monte Carlo studies showed
that  the  dijet   invariant  mass  resolution  for  signal   events is 16
GeV/c$^2$. The transverse momentum of the two gauge bosons was required to
be within   $|p_T^{jj}-p_T^{e\nu}| <  40$ GeV/c,  as  expected for $WW/WZ$
production.

There are two  major sources of  background to $WW/$  $WZ \rightarrow e\nu
jj$ production: (i) $W  + \geq 2$ jets with  $W\rightarrow e\nu$; and (ii)
QCD multijet events where one of the  jets is misidentified as an electron
and there is significant ${\hbox{$E$\kern-0.6em\lower-.1ex\hbox{/}}}_T$ in
the event due  to  mismeasurement. Other  backgrounds such  as: $t\bar{t}$
production   with   subsequent  decay  to    $W^+bW^-\bar{b}$  followed by
$W\rightarrow  e\nu$; $WW$ or $WZ$  production with  $W\rightarrow\tau\nu$
followed by  $\tau\rightarrow  e\nu\bar{\nu}$;  $ZX\rightarrow eeX$, where
one electron is mismeasured or not identified; and $ZX\rightarrow \tau\tau
X$ with $\tau\rightarrow e\nu\bar{\nu}$, are negligible.

The    $W+\geq  2$  jets    background  was    estimated  using  the  {\sc
vecbos}~\cite{vecbos} event generator, with $Q^2=(p_{T}^j)^2$, followed by
parton  fragmentation using the {\sc   herwig}~\cite{herwig} package and a
detailed  {\sc   geant}~\cite{geant}  based   simulation of the  detector.
Normalization   of  the  $W+\geq 2$  jets   background was   determined by
comparing the number of events  expected from the {\sc vecbos} estimate to
the number of  candidate events  outside the dijet  mass window, after the
multijet backgrounds had been  subtracted. The systematic uncertainties in
this background are due  to the normalization  and to the jet energy scale
correction.  The  multijet  background  was  estimated  following the same
procedure used in our previous analysis~\cite{two}. The background sample,
which consisted  of data events  containing a jet  satisfying the electron
trigger selection but failing the  electron identification, was normalized
to        the          signal         sample     in      the        region
${\hbox{$E$\kern-0.6em\lower-.1ex\hbox{/}}}_T  <  15$ GeV where the actual
$WW/WZ$  contribution is negligible.  The number of  background events was
then determined  from this scaled  background sample  with the rest of the
selection    criteria      applied~\cite{Thesis}.  The    backgrounds from
$t\bar{t}\rightarrow      W^+bW^-\bar{b}$  and  other  minor  sources were
estimated using the {\sc isajet} event generator~\cite{isajet} followed by
detector simulation.  Table I summarizes the  background estimates and the
total number of events seen. The  number of observed events was consistent
with the background estimates which dominate the SM $WW/WZ$ signal.

The  trigger  and  offline   electron   identification   efficiencies were
estimated  using   $Z\rightarrow ee$  events. The  trigger  efficiency was
($98.1 \pm  1.9$)\%~\cite{one}. The  electron  identification efficiencies
were  found  to be   ($74.5\pm  1.1$)\% in  the  central   calorimeter and
($61.9\pm 1.1$)\%  in the end  calorimeters. We  studied the $W\rightarrow
jj$  efficiency for the  jet cone  size  ${\cal{R}} = 0.5$  using the {\sc
isajet}  and {\sc    pythia}~\cite{pythia}  event  generators  followed by
detector simulation. The selection  criteria were applied to these samples
and it was found that  the efficiency was  $\approx 50$\% for $p_T^W< 250$
GeV/c and that this decreased  significantly for $p_T^W> 250$ GeV/c due to
merging of  the two  jets into one.  The  efficiencies  obtained from {\sc
isajet} were used to  estimate the detection  efficiencies of the $WW(WZ)$
processes since they gave more conservative results.

       	The overall  event selection  efficiency was  calculated using the
leading order event generator of Ref.~\cite{Zeep} to generate four-momenta
for $WW$  and  $WZ$  processes as a  function  of the  coupling  parameter
values. A  fast  detector  simulation was  used to  take into  account the
detector  resolutions and  efficiencies described  above. Higher order QCD
effects were approximated by a  $K$--factor $= 1 + \frac{8}{9} \pi\alpha_s
=  1.34$~\cite{Xsec}  and a  smearing of  the  transverse  momentum of the
diboson system according to the experimentally determined $p_T^Z$ spectrum
from  the   inclusive   $Z\rightarrow  ee$  sample.  The  total  selection
efficiencies for the  detection of SM $WW$ and  $WZ$ events were estimated
to be  [$  14.7\pm\   0.2$~(stat)~$\pm\   1.2$~(syst)]\%  and  [$14. 6\pm\
0.4$~(stat)~$\pm\ 1.1$~(syst)]\%, respectively. The systematic uncertainty
(8\%)   includes:  electron  trigger and   selection   efficiencies (1\%);
${\hbox{$E$\kern-0.6em\lower-.1ex\hbox{/}}}_T$   smearing and $p_T$ of the
$WW/WZ$ diboson system (5\%); difference between the {\sc isajet} and {\sc
pythia} programs for  $W\rightarrow jj$  efficiency parametrization (5\%);
statistical uncertainty  for $W\rightarrow jj$  efficiency parametrization
(2\%); and jet energy scale (3\%).

\begin{figure}[h]
\centerline{\psfig{file=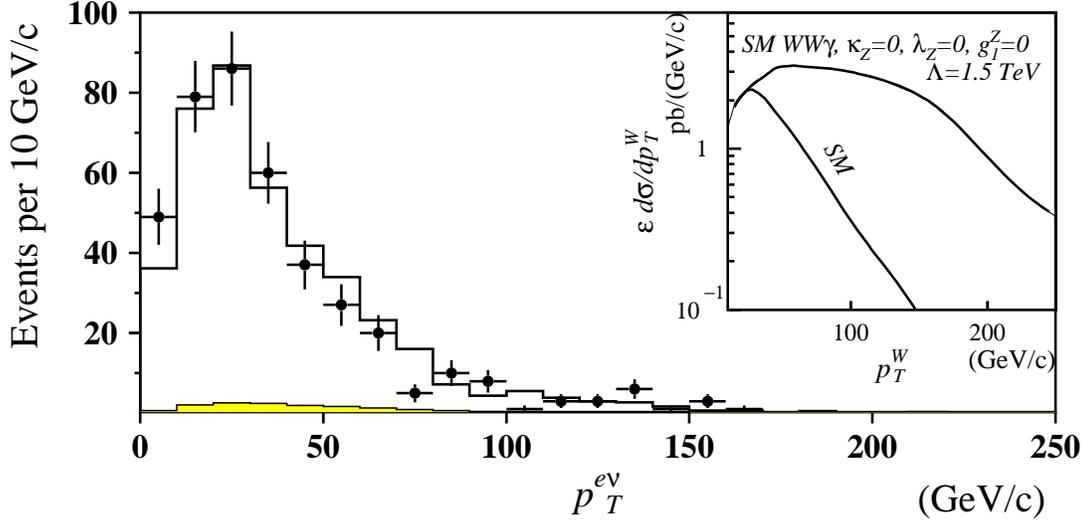,width=6.0in}} 
\caption{
$p_T$ distributions of the $e\nu$  system for the 1993--1995 data set. The
points with  error bars  represent  the data.  The solid  histogram is the
total background estimate plus the  SM Monte Carlo predictions of $WW$ and
$WZ$ production (shown as shaded histogram). The inset shows the predicted
$d\sigma/dp_T^W$,     folded  with the   detection   efficiencies,  for SM
$WW\gamma$ and  $WWZ$ couplings  (lower curve), and  for SM $WW\gamma$ and
the indicated anomalous $WWZ$ couplings (upper curve). } 
\label{fig:ptw_enu} 
\end{figure}

The expected  signal for  $WW$ plus $WZ$  production with  SM couplings is
$20.7\pm  3.2$ events  based on the  total  integrated  luminosity of 96.0
pb$^{-1}$.  Figure~\ref{fig:ptw_enu} shows the  $p_T^{e\nu}$ distributions
for candidate events from 1993--1995  data, total background estimate plus
SM expectations, and SM  expectations for $WW$  and $WZ$ production, after
all selection  criteria have  been applied.  There is no  clear difference
between  the   observed   $p_T^{e\nu}$  spectrum  and that   expected from
background plus SM $WW$ and $WZ$ prediction. 

Using the detection  efficiencies for SM $WW$  and $WZ$ production and the
background subtracted signal, and  assuming the SM ratio of cross sections
for $WW$ and $WZ$ production, we can  set an upper limit at the 95\% CL on
the cross section $\sigma(p\bar{p}\rightarrow W^+W^-X)$ of $76$ pb. 

Since we observed no excess of high  $p_T^{e\nu}$ events, large deviations
from  the SM   trilinear  coupling  values  are  excluded.   Limits on the
anomalous coupling  parameters were set by  performing a binned likelihood
fit to the  observed  $p_T^{e\nu}$  spectrum  with the Monte  Carlo signal
prediction plus the estimated  background. Unequal width bins were used to
evenly distribute the observed  events, especially those at the end of the
spectrum. In each  $p_T^{e\nu}$ bin for a given  set of anomalous coupling
parameters, we  calculated the  probability for the  sum of the background
estimate and Monte  Carlo $WW/WZ$ prediction to  fluctuate to the observed
number of events. The limits on the anomalous coupling parameters are from
a combined  likelihood  fit to both  data sets.  The  uncertainties in the
background  estimates,   efficiencies,  integrated  luminosity, and higher
order  QCD   corrections to  the  signal  were  convoluted  with  Gaussian
distributions into the  likelihood function.  Uncertainties common to both
analyses, e.g. theoretical uncertainties, were convoluted only once. 

\begin{figure}[h]
\centerline{\hbox{
\psfig{figure=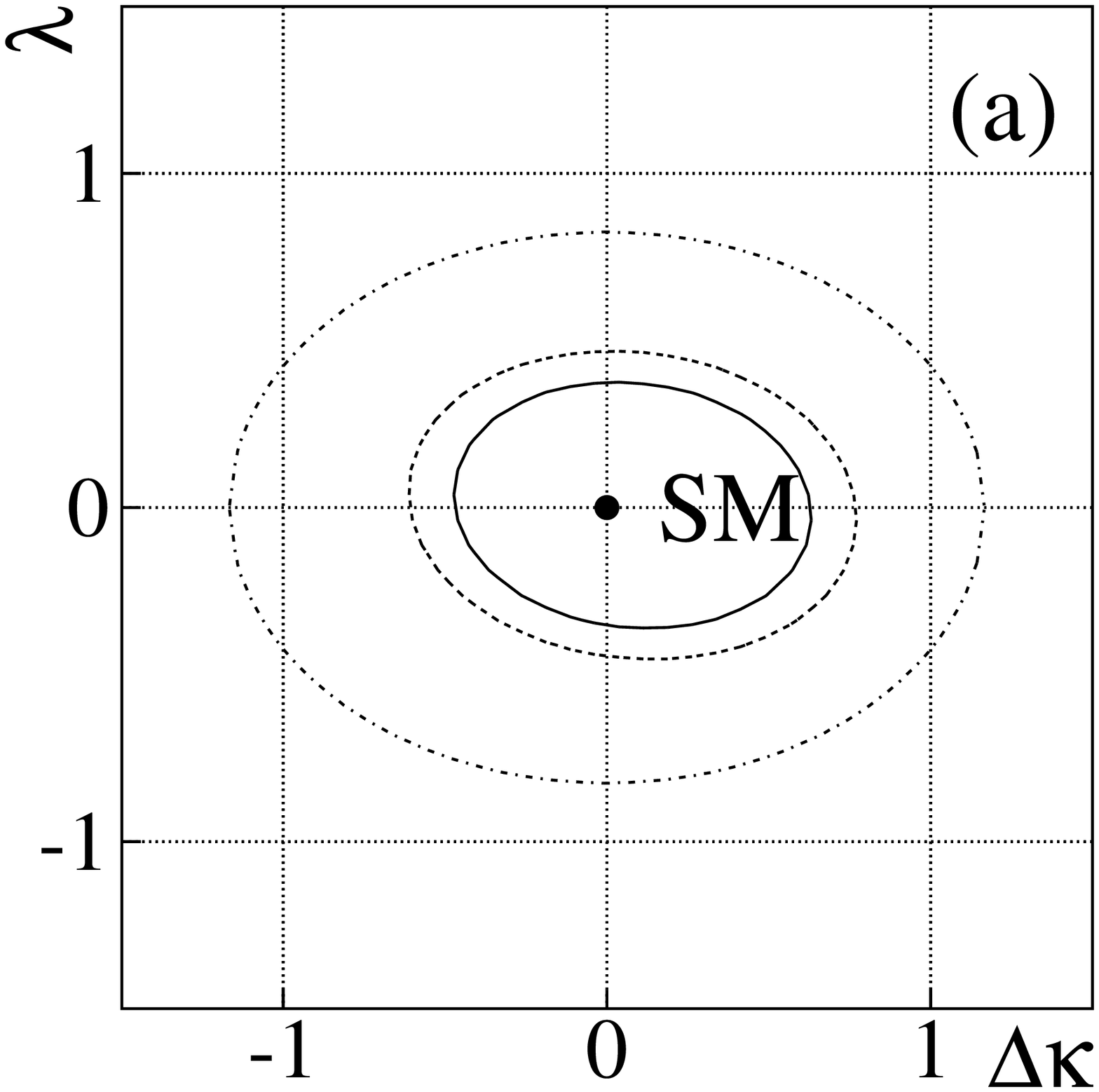,width=3.1in,height=3.0in}
\psfig{figure=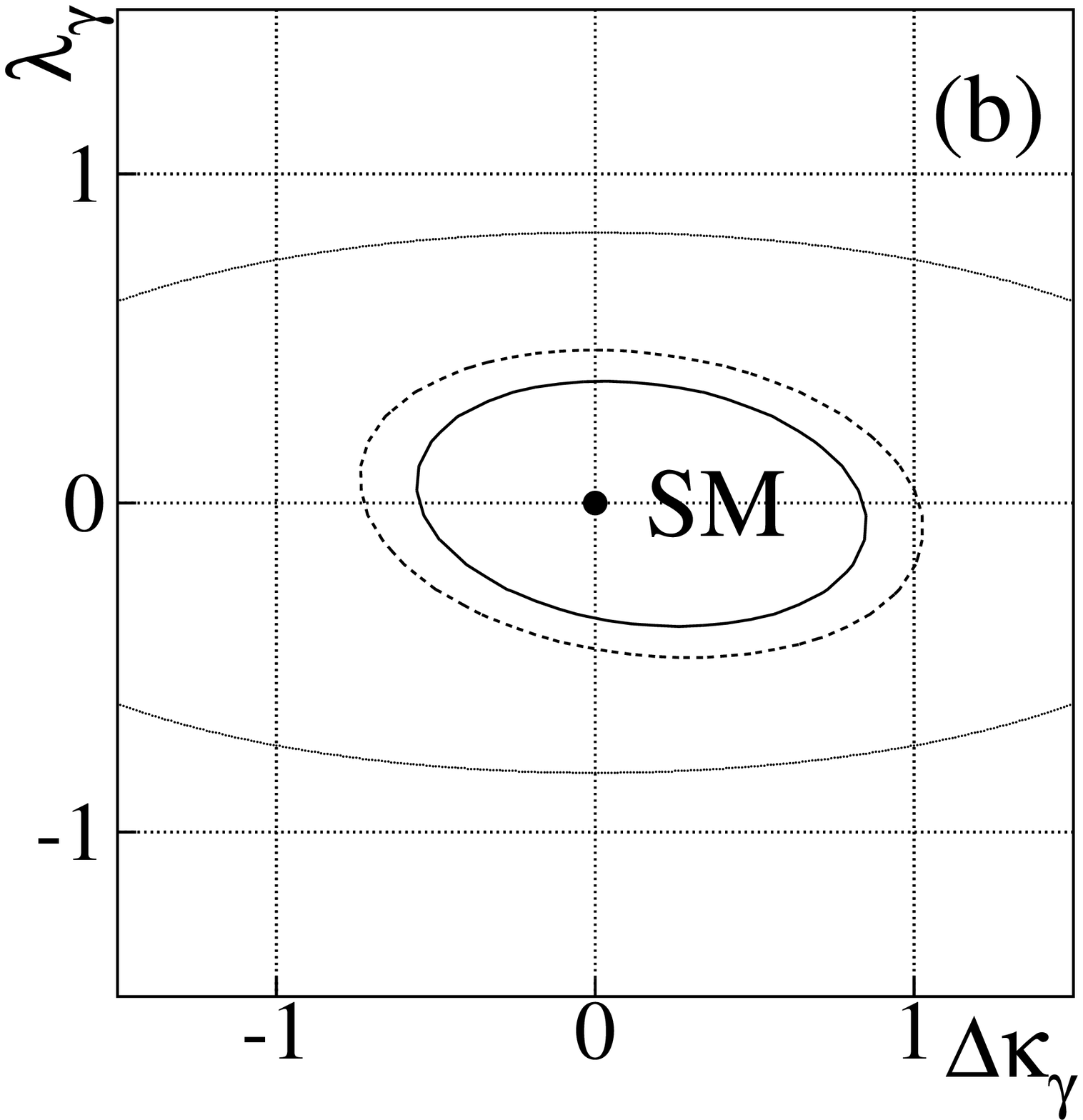,width=3.1in,height=3.0in}}}
\centerline{\hbox{
\psfig{figure=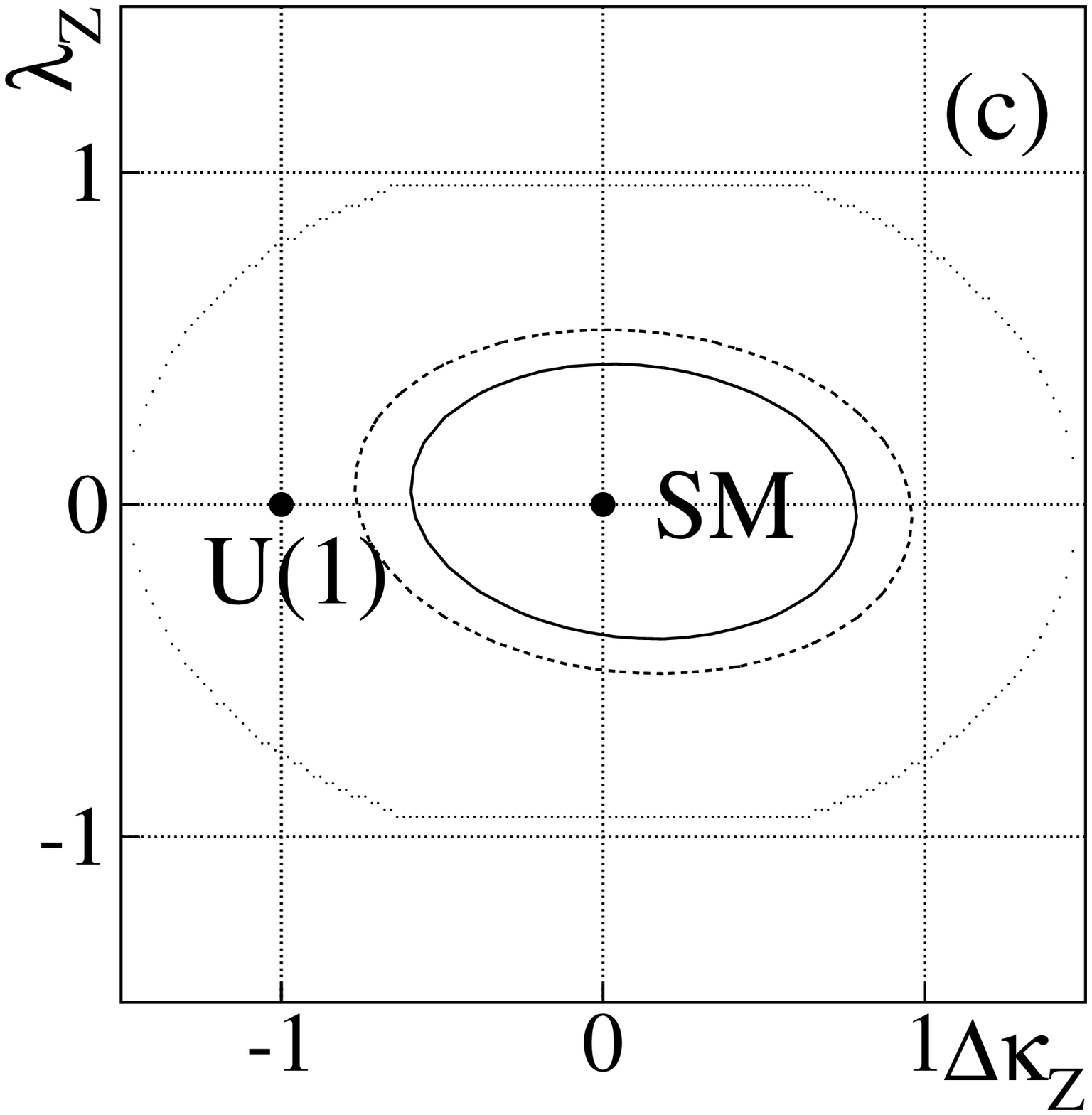,width=3.1in,height=3.0in}
\psfig{figure=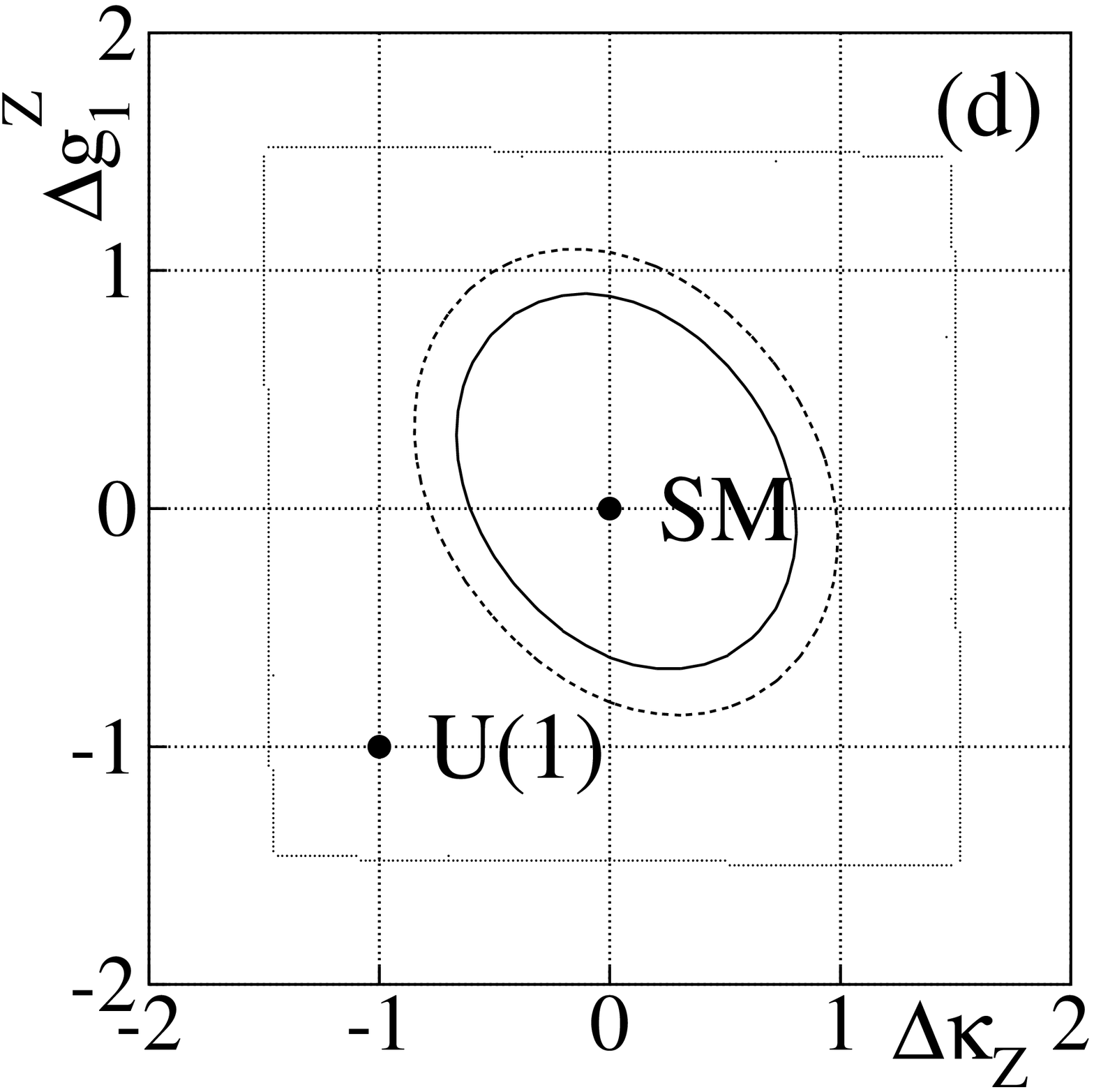,width=3.1in,height=3.0in}}}
\caption{
Limits   on    $CP$--conserving   anomalous   couplings    parameters: (a)
$\Delta\kappa\equiv\Delta\kappa_{\gamma}=\Delta\kappa_{Z}$,               
$\lambda\equiv\lambda_{\gamma}=\lambda_{Z}$;   (b) HISZ relations; (c) and
(d) SM $WW\gamma$ couplings. The inner and middle curves represent 95\% CL
one- and two-de\-gree-\-of-\-freedom exclusion contours, respectively. The
outermost  curves show  S--matrix unitarity  bounds.  $\Lambda=1.5$~TeV is
used for all four cases. The SM prediction is $\Delta\kappa=0, \lambda=0$,
$\Delta g_1^Z=0$.}
\label{fig:contours}
\end{figure}

In  Fig.~\ref{fig:contours}, bounds  on four pairs of  coupling parameters
are shown using $\Lambda = 1.5$ TeV.  In each case all other couplings are
fixed to  their SM  values.  The one- and    two-degree-of-freedom 95\% CL
contour limits (corresponding to  likelihood function values 1.92 and 3.00
units below  the  maximum,  respectively) are  shown as the  inner curves,
along with the S-matrix  unitarity limits, shown  as the outermost curves,
which are  obtained by  evaluating  all (i.e.  $WW$,  $W\gamma$, and $WZ$)
processes. Figure~2(a)  shows the contour limits  when coupling parameters
for  $WW\gamma$ are assumed  to be equal to  those for  $WWZ$. Figure~2(b)
shows contour limits assuming HISZ relations~\cite{HISZ}. In Fig. 2(c) and
2(d) SM $WW\gamma$ couplings are assumed and the coupling limits for $WWZ$
are shown. 

When SM $WW\gamma$ couplings are  assumed, the U(1) point ($\kappa_Z = 0$,
$\lambda_Z =0$,  $g_1^Z = 0$) is  excluded at the 99\%  CL. This is direct
evidence for the existence of the $WWZ$ couplings.

Table II  lists the  95\% CL  axis  limits for  three  different values of
$\Lambda$                    and        four                  assumptions:
(i)~$\Delta\kappa\equiv\Delta\kappa_{\gamma}=\Delta\kappa_{Z}$,           
$\lambda\equiv\lambda_{\gamma}=\lambda_{Z}$; (ii)~HISZ relations; (iii)~SM
$WW\gamma$ couplings; and (iv)~SM $WWZ$ couplings. The results with the SM
$WW\gamma$  assumption are  unique to  $WW/WZ$ production  since the $WWZ$
couplings  are  not  accessible with   $W\gamma$  production.  The results
indicate that this  analysis is more sensitive  to $WWZ$ couplings than to
$WW\gamma$ ones  as expected from  the larger overall  couplings for $WWZ$
than $WW\gamma$~\cite{Baur}. The  dependence of the coupling parameters on
$\Lambda$ is clearly seen. Tighter limits are obtained when a larger value
for $\Lambda$ is used. When SM $WWZ$  couplings are assumed, our limits on
$\lambda_{\gamma}$ and  $\Delta\kappa_{\gamma}$ with $\Lambda=2.0$ TeV are
not tight enough to lie within the S-matrix unitarity limit. 

In conclusion, we have presented  limits on anomalous $WWZ$ and $WW\gamma$
coupling   parameters  which  are the  most  stringent  to date.  They are
significantly tighter than those  from the analyses of the 1992--1993 data
set~\cite{CDF,two},   and   significantly  better on   $\Delta\kappa$ (but
comparable on $\lambda$) to those measured using $W\gamma$ production with
the 1992--1995 data set~\cite{one}.

We thank U. Baur for useful discussions and D. Zeppenfeld for providing us
with the  $WW$ and $WZ$  Monte Carlo   generators. We thank  the staffs at
Fermilab and  collaborating  institutions for their  contributions to this
work, and acknowledge support from  the  Department of Energy and National
Science   Foundation  (U.S.A.),    Commissariat  \` a  L'Energie  Atomique
(France),   State Committee  for Science  and Technology  and Ministry for
Atomic Energy   (Russia), CNPq (Brazil),  Departments of Atomic Energy and
Science and Education  (India), Colciencias  (Colombia), CONACyT (Mexico),
Ministry of Education  and KOSEF (Korea),  CONICET and UBACyT (Argentina),
and the A.P. Sloan Foundation.

\begin{table}[h]
\caption{Summary of signal and backgrounds.}
\begin{tabular}{l r @{$\pm$} l r @{$\pm$} l}
                       &\multicolumn{2}{c}{1992--1993}
                       &\multicolumn{2}{c}{1993--1995}         \\ 
\hline
Luminosity             &\multicolumn{2}{c}{13.7 pb$^{-1}$}
                       &\multicolumn{2}{c}{82.3 pb$^{-1}$} \\ 
Backgrounds            &\multicolumn{2}{c}{}&\multicolumn{2}{c}{}\\ 
~~$W + \geq $ 2 jets              & 62.2 & 13.0 & 279.5 & 36.0 \\
~~QCD Multijet                    & 12.2 &  2.6 & 104.3 & 12.3 \\
~~$t\bar{t}\rightarrow e\nu\ jjX$ &  0.9 &  0.1 &   3.7 &  1.3 \\
Total Background                  & 75.3 & 13.3 & 387.5 & 38.1 \\
Data                   &\multicolumn{2}{c}{84}&\multicolumn{2}{c}{399}\\
SM $WW$+$WZ$ prediction           &  3.2 &  0.6 &  17.5 &  3.0 \\
\end{tabular}
\label{table:1a_1b}
\end{table}	

\begin{table}[h]
\caption{Axis  limits at the  95\% CL with  various  assumptions and three
different $\Lambda$ values.}
\begin{tabular}{lc|c|c|c}
\multicolumn{2}{c|}{Couplings / $\Lambda$(TeV)} & 1.0 & 1.5 & 2.0\\
\hline
(i)&$\lambda_{\gamma}= \lambda_Z$ & $-0.42$, $0.45$ & $-0.36$, $0.39$ &
    $-0.33$, $0.36$\\
   &$\Delta\kappa_{\gamma}=\Delta\kappa_Z$ & $-0.55$, $0.79$ & 
    $-0.47$, $0.63$ & $-0.43 $, $0.59$\\
(ii)&$\lambda_{\gamma}$ (HISZ)       & $-0.42$, $0.45$ & $-0.36$, $0.38$ &
    $-0.34$, $0.36$\\
   &$\Delta\kappa_{\gamma}$ (HISZ)  & $-0.69$, $1.04$ & $-0.56$, $0.85$ &
    $-0.53$, $0.78$\\
(iii)&$\lambda_Z$ (SM $WW\gamma$)     & $-0.47$, $0.51$ & $-0.40$, $0.43$ &
    $-0.37$, $0.40$\\
   &$\Delta\kappa_Z$ (SM $WW\gamma$)& $-0.74$, $0.99$ & $-0.60$, $0.79$ &
    $-0.54$, $0.72$\\
   &$\Delta g_1^Z$  (SM $WW\gamma$) & $-0.75$, $1.06$ & $-0.64$, $0.89$ &
    $-0.60$, $0.81$\\ 
(iv)&$\lambda_{\gamma}$ (SM $WWZ$)   & $-1.28$, $1.33$ & $-1.21$, $1.25$ &
    \\
   &$\Delta\kappa_\gamma$ (SM $WWZ$)& $-1.60$, $2.03$ & $-1.38$, $1.70$ &
    \\
\end{tabular}
\label{table:lim_com}
\end{table}

\newpage

\newpage

\begin{references}

%
\bibitem[*]{beijing}
Visitor from IHEP, Beijing, China.

\bibitem[\dag]{ecuador}
Visitor from Universidad San Francisco de Quito, Quito, Ecuador.

\vskip 0.25cm

\bibitem{UA2} J.~Alitti {\it et al.\/} (UA2 Collaboration), Phys.
Lett. {\bf B277}, 194 (1992).

\bibitem{CDF} F.~Abe {\it et al.\/} (CDF Collaboration), Phys. Rev.
Lett. {\bf 74}, 1936 (1995); 
{\it ibid.\/}, {\bf 74}, 1941 (1995);
{\it ibid.\/}, {\bf 75}, 1017 (1995);
F.~Abe {\it et al.\/} (CDF Collaboration), Fermilab--Pub--96/311--E,
to be published in Phys. Rev. Lett.

\bibitem{one} S.~Abachi {\it et al.\/} (D\O\ Collaboration), 
Phys. Rev. Lett. {\bf 75}, 1023 (1995);
{\it ibid.\/}, {\bf 75}, 1028 (1995);
{\it ibid.\/}, {\bf 75}, 1034 (1995);
{\it ibid.\/}, {\bf 78}, 3634 (1997);
{\it ibid.\/}, {\bf 78}, 3640 (1997).

\bibitem{two} S.~Abachi {\it et al.\/} (D\O\ Collaboration), Phys. Rev.
Lett. {\bf 77}, 3303 (1996).

\bibitem{TopPRD} S.~Abachi  {\it et al.} (D\O\ Collaboration), 
Fermilab--Pub--97/088--E, hep--ex/9704004, submitted to  Phys. Rev. D.

\bibitem{Baur} K.~Hagiwara, R.D.~Peccei, D.~Zeppenfeld and K.~Hikasa,
Nucl. Phys. {\bf B282}, 253 (1987).

\bibitem{Xsec} J.~Ohnemus, Phys. Rev. D {\bf 44}, 1403 (1991); 
{\it ibid.\/}, {\bf 44}, 3477 (1991). 

\bibitem{Detector} S.~Abachi {\it et al.\/} (D\O\ Collaboration), Nucl. 
Instrum.  Methods {\bf A338}, 185 (1994). 

\bibitem{Thesis} A.~S{\'a}n\-chez-Her\-n{\'a}n\-dez, Ph.D.
Dissertation, CINVES\-TAV,
Me\-xi\-co Ci\-ty, Me\-xi\-co (1997), un\-published.

\bibitem{squarks} S.~Abachi {\it et al.\/} (D\O\ Collaboration), Phys. Rev.
Lett. {\bf 75}, 618 (1995).

\bibitem{Scale} R.~Kehoe (for the D\O\ Collaboration), preprint 
Fermilab--Conf--96/284--E, to appear in Proc. 6th International Conf. 
on Calorimetry in High Energy Physics, Frascati (1996).

\bibitem{vecbos} F.A.~Berends {\it et al.\/}, Nucl. Phys. {\bf B357},
32 (1991). We used version 3.0. 

\bibitem{herwig} G.~Marchesini {\it  et al.\/}, Comput. Phys. Commun. {\bf
67}, 465 (1992). We used version 5.7.

\bibitem{geant}    F.~Carminati  {\it et  al.\/}, {\it  {\sc  geant} Users
Guide,} CERN Program Library Long Writeup WS013 (1993), unpublished. 
 
\bibitem{isajet} F.~Paige and S.~Protopopescu, BNL Report BNL38034 (1986),
unpublished. We used version 7.22.

\bibitem{pythia} T.~Sj\"ostrand, Comput. Phys. Commun. {\bf 82}, 74
(1994).

\bibitem{Zeep}  K.~Hagiwara,  J.~Woodside and  D.~Zeppenfeld, Phys. Rev. D
{\bf 41}, 2113 (1990); D.~Zeppenfeld (private communication).

\bibitem{HISZ}  K.~Hagiwara, S.~Ishihara,  R.~Szalapski and D.~Zeppenfeld,
Phys. Rev. D {\bf 48},  2182 (1993); Phys. Lett.  B {\bf 283}, 353 (1992).
They parametrize the $WWZ$ couplings in terms of the $WW\gamma$ couplings:
$\Delta\kappa_Z =  \Delta\kappa_{\gamma} (1 - \tan^2 \theta_W)/2$, $\Delta
g_1^Z  =       \Delta\kappa_{\gamma}/2\cos^2    \theta_W$ and   $\lambda_Z
=\lambda_{\gamma}$.

\end{references}
\end{document}